\documentclass[final,1p,times]{elsarticle}
\usepackage{graphicx}
\usepackage{amssymb}
\usepackage{amsmath}
\usepackage{mathtools}
\usepackage{bm}
\usepackage[hyphens]{url}
\usepackage{hyperref}
\usepackage{algorithm}
\usepackage{algpseudocode}
\usepackage{lineno}
\usepackage{booktabs}
\usepackage{adjustbox}
\usepackage{subcaption}
\usepackage{lipsum}
\journal{Journal of the Royal Society Interface}

\begin{document}

\begin{frontmatter}


\title{Uncertainty quantification of a 3D In-Stent Restenosis model with surrogate modelling}



\author[uvaaddress]{Dongwei Ye\corref{mycorrespondingauthor}}
\cortext[mycorrespondingauthor]{Corresponding authors}
\ead{d.ye@uva.nl}

\author[uvaaddress,itmoaddress]{Pavel Zun}
\ead{p.zun@uva.nl}

\author[uvaaddress]{Valeria Krzhizhanovskaya}
\ead{V.Krzhizhanovskaya@uva.nl}

\author[uvaaddress]{Alfons~G.~Hoekstra}
\ead{A.G.Hoekstra@uva.nl}

\address[uvaaddress]{Computational Science Lab, Institute for Informatics, Faculty of Science, University of Amsterdam, The Netherlands}

\address[itmoaddress]{National Center for Cognitive Research, ITMO University, Saint Petersburg, Russia}

\begin{abstract}
In-Stent Restenosis is a recurrence of coronary artery narrowing due to vascular injury caused by balloon dilation and stent placement. It may lead to the relapse of angina symptoms or to an acute coronary syndrome. An uncertainty quantification of a model for In-Stent Restenosis with four uncertain parameters (endothelium regeneration time, the threshold strain for smooth muscle cells bond breaking, blood flow velocity and the percentage of fenestration in the internal elastic lamina) is presented. Two quantities of interest were studied, namely the average cross-sectional area and the maximum relative area loss in a vessel. Due to the computational intensity of the model and the number of evaluations required for the uncertainty quantification, a surrogate model, based on Gaussian process regression with proper orthogonal decomposition, was developed which subsequently replaced the original In-Stent Restenosis model in the uncertainty quantification. A detailed analysis of the uncertainty propagation and sensitivity analysis is presented. Around 11\% and 16\% of uncertainty are observed on the average cross-sectional area and maximum relative area loss respectively, and the uncertainty estimates shows that a higher fenestration mainly determines uncertainty in the neointimal growth at the initial stage of the process. On the other hand, the uncertainty in blood flow velocity and endothelium regeneration time mainly determine the uncertainty in the quantities of interest at the later, clinically relevant stages of the restenosis process. The uncertainty in the threshold strain is relatively small compared to the other uncertain parameters.

\end{abstract}


\end{frontmatter}

% \linenumbers
\numberwithin{table}{section}
\section{Introduction}\label{Sec:intro}
Coronary heart disease is mainly due to the accumulation and development of atherosclerotic plaque which narrows the vessel lumen and reduces the flow of blood. It can cause ischemia or further evolve into a myocardial infarction. The most common treatments is percutaneous coronary intervention with stent deployment \cite{PTCA1979,stent2011}. However, the balloon dilation for stent placement not only moves out of the way the plaque blocking the blood flow, but also denudes the endothelium layer and damages the vessel wall. It then triggers smooth muscle cells activation, proliferation and migration, and extracellular matrix formation, as well as other processes, e.g. inflammation and platelet aggregation \cite{Jukema2012PC1,Jukema2012PC2}. This may lead to excessive growth of neointima, a condition known as In-Stent Restenosis (ISR).

To study the mechanism of restenosis, a multiscale model for ISR was proposed \cite{Evans2008} and a first two-dimensional version of that model (coined ISR2D) was developed and studied in detail \cite{Tahir2011,Tahir2013,Tahir2014}. The model consists of three submodels, an initial condition model, an agent-based smooth muscle cell (SMC) model, and a blood flow model. It has been applied to investigate the effect of functional endothelium regeneration and the impact of stent deployment and design on restenosis \cite{Tahir2011,Tahir2013,Nikishova2018,Ye2020}. Most recently, the effects of local blood flow dynamics with scenarios of adaptive and non-adaptive coronary vasculature on restenosis was studied based on the ISR2D model \cite{ZUN2021110361}. The two-dimensional model is however a simplification of the actual pathology. Therefore, a more comprehensive three dimensional model (coined ISR3D) was developed and compared to in-vivo experimental scenarios \cite{ISR3D2017,Zun2019}. 

Uncertainty quantification (UQ) is widely applied to study the effect of uncertainties in initial or boundary conditions and of other parameters of computational models on their simulated quantities of interest. Common uncertainty quantification methods, such as those based on Monte Carlo method \cite{caflisch1998,Heinrich2001,Peherstorfer2016}, polynomial chaos expansion \cite{CRESTAUX20091161,Gratiet2016} and stochastic collocation \cite{Sankaran2011,Jakeman2013} require a large number of simulations to provide enough data for the numerical integration of the statistical estimator \cite{fang2005design}. However it might be prohibitive for computationally expensive models, such as ISR3D, to achieve this. One solution could be to adopt surrogate modelling, by which a surrogate model (or metamodel) is developed to approximate the response of the original model at a relatively low cost. Subsequently, this surrogate model replaces the original simulation to realise the evaluations required for the UQ.

The construction of a surrogate model can be categorized into three types: simplified models, projection-based methods and data-fit methods \cite{Survey2018}. Simplified models refer to a rough approximation based on simplifications of the simulated system such as spatial dimensionality reduction \cite{Koepp2018,Coccarelli2021} or coarse-grid discretisations \cite{ZHANG2021113485,SEN2018434}. The projection-based methods proceed by identifying a low-dimensional subspace that is constructed to retain the essential character of the system input-output mapping. One state-of-the-art projection-based methods is Proper Orthogonal Decomposition (POD) \cite{chatterjee2000introduction,Berkooz1993,Guo2018}. It captures the dominant components of a high-dimensional process with low-dimensional approximations. Finally, the data-fit methods map out latent functions between input and output. Common methods of this type of surrogates are support vector machines \cite{wang2005support}, neural networks \cite{goodfellow2016deep} or Gaussian processes \cite{Rasmussen2004}. 
 
Gaussian process (GP) regression is widely applied in uncertainty estimation and reliability analysis due to its non-parametric and Bayesian inference nature \cite{MARREL2009742,MARREL20084731,Muehlenstaedt2012,Le2014}. It was first proposed by Krige for geostatistics \cite{Krige1951}, and later extensively studied and extended to solve the regression problem under different scenarios, such as multi-task/multi-output Gaussian process for vector-valued function \cite{LIU2018102}, heteroscedastic Gaussian process for input dependent noise scenarios \cite{Quoc2005,wrro136813,BILIONIS20125718}, sparse Gaussian process with inducing inputs for efficient training of large dataset \cite{Matthias2016,NIPS2005_4491777b} or deep Gaussian process with a hierarchical structure to capture more complex processes \cite{damianou13a}. 

Generally GPs are designed for a scalar output and become cumbersome when multi-output is required due to the large kernel used for coregionalization. The complexity of multi-output Gaussian process (MOGP) is associated with the dimension of output and the number of training samples. The computational cost of MOGP can easily become prohibitively expensive if the desired output dimension is high. One alternative solution is to apply dimensionality reduction techniques, such as Proper orthogonal decomposition \cite{chatterjee2000introduction}, to the model response before regression. The regression prediction is hence no longer the model response but the projection coefficients of the response. Due to the limited amount of projection coefficients required for the reconstruction of the output space, several single-output GPs are sufficient in this case. This method has been widely applied for time-dependent problems \cite{GUO201975,MAULIK2021132797}, computational fluid dynamics \cite{YANG2020108596}, etc.

Here, the uncertainty propagation due to four uncertain parameters of the ISR3D model (endothelium regeneration time, the threshold strain for smooth muscle cells bond breaking, blood flow velocity and the percentage of fenestration in the internal elastic lamina) is investigated. The Quantities of Interest (QoIs) are the average cross-sectional area of the lumen and the maximum relative area loss as a function time. We applied POD to reduce the dimension of the output and used Gaussian process regression as the surrogate model to map the uncertain inputs to the projection coefficients of the POD. With this computationally efficient surrogate model, uncertainty estimations and sensitivity analysis of the restenosis process are conducted and analysed.   

The paper is arranged as follows. The details of the ISR3D model are introduced in Section~\ref{sec:ISR3D}. The construction of the surrogate model with POD and GP is described in Section~\ref{sec:surrogate}. The uncertain parameters and uncertainty estimations are presented in Section~\ref{sec:UQ}. The results of uncertainty estimates and sensitivity analysis are presented in Section~\ref{sec:result} followed by a discussion in Section~\ref{sec:dis} and the conclusions in Section~\ref{sec:con}. 

\section{In-Stent Restenosis 3D Model}\label{sec:ISR3D}

\begin{figure}[tb]
\centering
 \includegraphics[width=1.\textwidth]{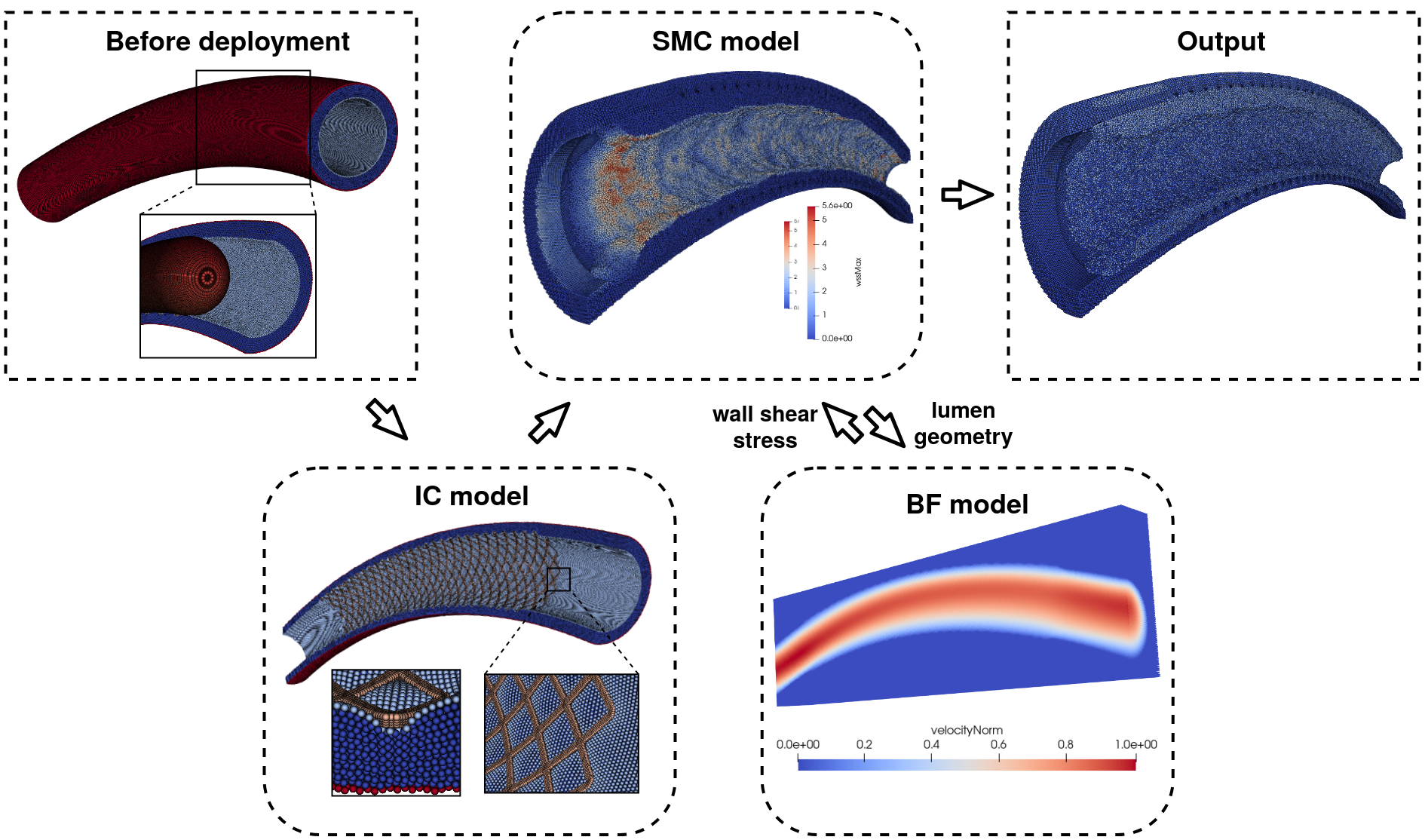}
 \caption{A schematic diagram of the ISR3D modeL. After the initial deployment of the stent with the IC model, the SMC and BF models run iteratively until the end of the simulation. At every timestep the SMC model passes the current lumen geometry to the BF model which then updates the blood flow and send the wall shear stress back to the SMC model which then computes lumen growth, based on the wall shear stress.}
\label{fig:ISR3D}
\end{figure}

In-Stent Restenosis 3D (ISR3D) is a multiscale computational model simulating the post-stenting neointima growth in a coronary artery \cite{ISR3D2017,Zun2019}. It mainly consists of three single-scale submodels: the Initial Condition (IC) model, the Smooth Muscle Cells (SMC) model (including details of the vascular wall, such as lamina and the endothelial cells) and the Blood Flow (BF) model. A schematic diagram of ISR3D is shown in Figure~\ref{fig:ISR3D}. 

The SMC model has two parts, one deals with the biomechanics of the vessel wall post-stenting, while the other deals with the SMC biology, mainly in relation to proliferation and production of extracellular matrix. The mechanical part of the SMC model simulates the mechanical response of the vessel wall, based on cell-cell pairwise repulsive and attractive forces and calculating the cell displacements. Each SMC of the vessel wall is modelled as a spherical agent, and the interactions between them are provided by potential and bond forces. The effective radii of particles represents the radii of corresponding cells and changes during the growth governed by the biological solver \cite{Zun2019}. 

The biological model of SMCs describes the cell cycle dynamics. Cell lifecycle is a sequence of growth, replication and division of the cell; at the end of the lifecycle, the cell divides into two daughter cells. The processes that influence the cell lifecycle take place in the 30 $\mu m$ neighbourhood around the cell; the time scale of one cycle is around 24-48 hours.

The growth of separate cells is modelled by a finite-state automaton. Each cell can be in a state of growth (G1), synthesis/secondary growth/mitosis (S/G2/M), or a quiescent state (G0). Cells evolve from one state to the next, and stop or die under the influence of external factors such as contact inhibition (the mechanical stresses in between SMCs) or the concentration of nitric oxide. The biological model provides new radii, states of the cells as its output, and also the initial coordinates for newly formed cells. Growth of the neointima takes several dozens of cell cycles and stops several weeks after the stenting procedure \cite{Zun2019}.

The BF model is a pressure-driven fluid dynamics model, which provides relevant ranges of shear stresses on the vessel walls. The solver receives the lumen geometry every timestep from the SMC solver, simulates the steady-state blood flow and returns the wall shear stress information to the SMC model. The blood is assumed to be incompressible and Newtonian, and is modelled by the Lattice Boltzmann Method (LBM) \cite{kruger2017lattice} in 3D rectangular mesh (D3Q19). The inlet boundary condition for velocity is set to a parabolic profile and its maximum velocity is defined as one of the uncertain parameters. A Dirichlet pressure boundary condition is assigned at the outlet and the vessel wall is defined as a non-slip condition. The simulation is implemented with \textit{Palabos} \cite{LATT2021334}.

The initial stent deployment is performed by the initial condition (IC) model. The stent is expanded radially with a capsule-shaped balloon until it reaches a predefined deployment depth. As there is no uncertain input of the UQ experiment related to the IC model and all the simulations start from exactly the same post-deployment state, we exclude the IC model from the execution of the UQ. For further details about the ISR3D, see \cite{ISR3D2017,Zun2019}. A public version of the ISR3D model, which is studied in this paper, can be found on Github\footnote{https://github.com/ISR3D/ISR3D}.

\begin{figure}[tb]
\centering
 \includegraphics[width=1.\textwidth]{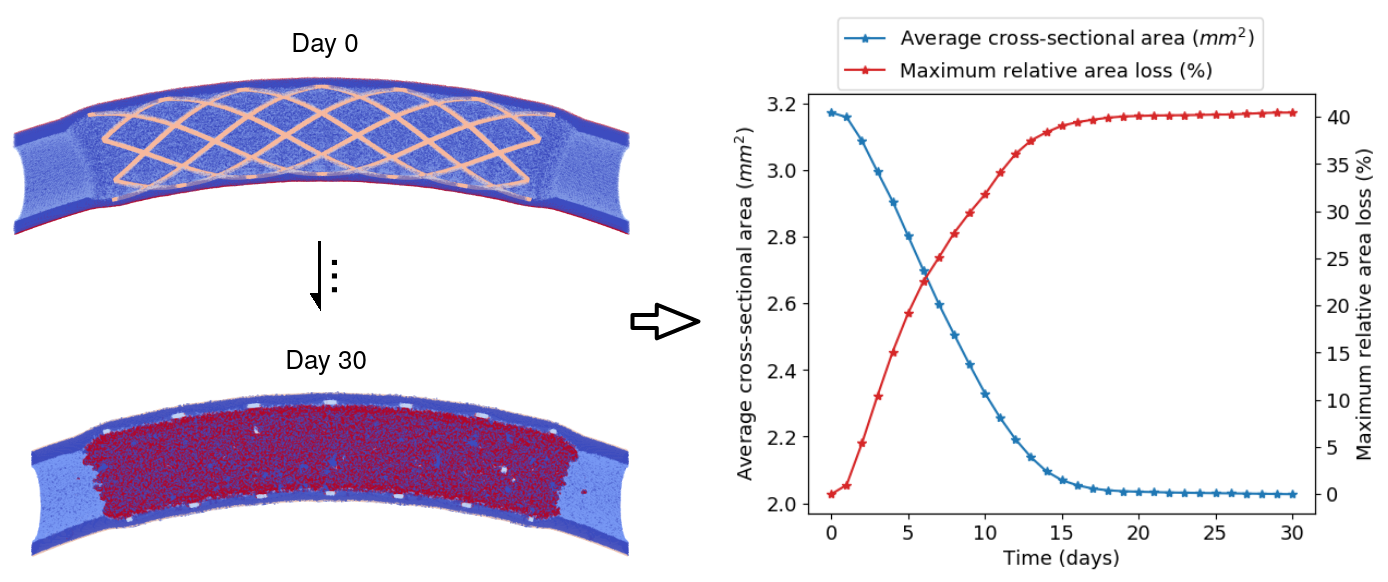}
 \caption{The simulation outcomes of ISR3D model (left) and corresponding QoIs measured over time (right). On the left-hand side, the blue part denotes the vessel wall, the beige part is the stent and the red part denotes the neointima. On the right-hand side, the average cross-sectional area and maximum relative area loss of the vessel at each day are measured and computed.}
\label{fig:Output}
\end{figure}

In the UQ experiments described here, the scenario of stenting a small porcine coronary vessel with $2~mm$ diameter is simulated. The entire length of the vessel is set to be $18~mm$ with a tunica width of $0.35~mm$ thickness and $1~mm$ lumen radius. The entire vessel is assumed to be slightly curved to obtain a more realistic blood flow pattern in the vessel. The stent applied in the simulations is made of intersecting spiral elements (shown in Figure~\ref{fig:Output}a left). It can be viewed as a simplified version of the NIR stent \cite{Stoeckel2002} and the deployment depth is set to be $0.25~mm$. The model is set to simulate the restenosis process up to 30 days after stenting.

The computational cost of ISR3D with a vessel and a stent of this size is rather expensive. A single run of the ISR3D simulation takes 500 to 600 core hours on a supercomputer node (a node with $2\times12$-core 2.6 GHz Intel Xeon E5-2690 v3 CPUs), depending on the total amount of neointima growth. For non-intrusive UQ methods, a large number of evaluations of the model are required for the statistical analysis and this becomes impractical for such a computationally intensive model. Therefore, to perform the UQ efficiently, a data-driven surrogate model based on GP and POD is developed to learn the latent function between the uncertain inputs and the QoIs, and applied to evaluate the model response in the UQ. 

Two QoIs are measured in the UQ experiment: the average cross-sectional area of the vessel lumen, and the maximum relative area loss. The lumen cross-sectional areas along the centerline of the vessel are obtained using an open-source toolkit VMTK\footnote{http://www.vmtk.org/}. The average value of this area over the considered vessel model at each timestep are used to evaluate how the uncertain parameters influence the total amount of neointima growth over time (shown in Figure~\ref{fig:Output} right). The relative area loss of the vessel shows the relative amount of neointima growth compared to the initial post-stenting cross-sectional area. Clinically, the restenosis is defined as the renarrowing of the lumen to more than 50\% occlusion \cite{Jukema2012PC1}. The maximum value of relative area loss of a vessel offers us a criterion to judge whether the restenosis happens or not. Note that both QoIs are evaluated as a function of time. The values at consecutive timesteps are highly correlated to each other.  

% To measure the neointimal growth, the cross-sectional areas along the centerline of the vessel are computed via an open source toolkit VMTK\footnote{http://www.vmtk.org/}. One of the examples is shown in Figure~\ref{}. 

% \begin{table}[tb]\label{tab:para}
% \centering
% \begin{tabular}{c|c|c|c|c|c}
% \toprule
% \textbf{Parameters}                                                         & \textbf{Value} & \textbf{Unit} & \textbf{Parameters} & \textbf{Value} & \textbf{Unit} \\ \midrule
% vessel length                                                               & 18             & mm            & stent length        & 12             & mm           \\
% tunica width                                                                & 0.175          & mm            & strut height        & 0.09           & mm           \\
% curvature of vessel                                                         & 1/4            & /             & strut width         & 0.09           & mm           \\
% curvature of  stent                                                         & 1/4            & /             & helix pitch         & 12             & ?            \\
% \begin{tabular}[c]{@{}c@{}}lumen radius\end{tabular} & 1              & mm            & number of struts    & 12             & /            \\ \bottomrule
% \end{tabular}
% \caption{Parameters of ISR3D simulations}
% \end{table}

\section{Surrogate modelling}\label{sec:surrogate}
\subsection{Proper orthogonal decomposition on model response}
Assume the response of the model is a series of responses (here, average cross-sectional areas of the lumen) over time $\bm{y} \in \mathbb{R}^{N_t}$, where $N_t$ is the dimension of the output vector. The proper orthogonal decomposition method can be applied to approximate the model responses by projecting the response to a low-rank space. The POD can be realized in three schemes, Karhunen-Loeve decomposition, principal component analysis, and the Singular Value Decomposition (SVD). In this work, the SVD method is applied for the decomposition \cite{Liang2002}. 

Consider a snapshot matrix $ \bm{S}\in \mathbb{R}^{N_t \times N_s}$ consisting of $N_s$ number of the model responses $\{\bm{y_1},\bm{y_2},...,\bm{y_{N_s}}\}$:
\begin{equation}
    \bm{S} = [\bm{y_1}|\bm{y_2}|...|\bm{y_{N_s}}], \hspace{.2cm} \text{where} \hspace{.1cm} N_t \gg N_s.
\end{equation}
The snapshot matrix can be decomposed into three matrices using singular value decomposition:
\begin{equation}
    \bm{S} =  \textbf{\textrm{U}} \bm{\Sigma} \textbf{\textrm{V}}^{\textrm{T}},
\end{equation}
where $\textbf{\textrm{U}}$ and $\textbf{\textrm{V}}$ denote left and right orthonormal matrices. $\bm{\Sigma}$ denotes a diagonal matrix with singular values $\sigma_i,  \text{where} \hspace{0.1cm} i = 1,...,N_s$ and $\sigma_1 \geq \sigma_2\geq ... \geq \sigma_{N_s} \geq 0$. 
% The $i$th basis vector $\bm{u_i}$ is taken from $i$th column of $\textbf{\textrm{U}}$ and associated to the singular value $\sigma_i$. 

The objective of POD is to find out a set of orthogonal basis $\bm{\Tilde{\Phi}} = \{ \phi_1, \phi_2, \cdots, \phi_k \}$  from the space $\mathcal{L} = \{ \bm{\Phi} \in \mathbb{R}^{N_t \times k} : \bm{\Phi}^\mathrm{T} \bm{\Phi} =\textbf{\textrm{I}} \}$  containing all possible orthogonal bases, such that the error introduced by the projection to low dimensional space could be minimized:
\begin{equation}
    \min_{\bm{\Phi} \in \mathcal{L}  } \sum_{i=1}^{N_s} \lVert \bm{y}_i - \bm{\Phi} \bm{\Phi}^T \bm{y}_i \rVert ^2 _{L^2} .
\end{equation}
By the Eckart-Young theorem \cite{Eckart1936}, the orthogonal basis with the basis vectors $\{\bm{u}_i\}_{i=1}^{k}$ taken
from the $i$th column of $\textbf{\textrm{U}}$ is the solution to such optimization problem. The relative energy captured by the projection to such low dimensional space consisting of the first $k$ columns of \textbf{\textrm{U}} can be evaluated by \cite{Anindya2000}:
\begin{equation}
   R_{\text{en}} = 1 - \frac{\sum^{N_s}_{i=1} \lVert \bm{y}_i - \bm{\Tilde{\Phi}} \bm{\Tilde{\Phi}}^T \bm{y}_i \rVert ^2 _{L^2}  }{\sum^{N_s}_{i=1} \lVert \bm{y}_i \rVert ^2 _{L^2}} = \frac{\sum^{k}_{i=1} \sigma_i^2}{\sum^{N_s}_{i=1} \sigma_i^2}
\end{equation}
We assume that if the relative energy $R_{\text{en}}$ is higher than 99.9\%, the approximation reconstructed by the first $k$ bases performs well enough. Since the values of $\sigma_i$ decay rapidly, a small $k$ would be sufficient to achieve the relative energy threshold. Once the basis vectors are obtained, any model response can be approximated by: $\bm{y} \approx \hat{\bm{y}}  = \sum^k_{i=1} \alpha_i \phi_i$, where $\alpha_i$ are the projection coefficients.

% In practice, such optimization problem could be also viewed as an eigenvalue problem $\bm{S}^\text{T}\bm{S} z_j = \lambda_j z_j$, the eigenvectors and eigenvalues of which are identical to the singular vectors and the square root of singular values of $\bm{S}$ respectively. 

% Once the basis vector is obtained, the response is approximated by: $\bm{y} \approx \bm{\hat{y}} = \sum^{k^*}_{i=1} \alpha_i \bm{u_i}$,
% where $\alpha_i$ are the projection coefficients. . 

\subsection{Gaussian process regression}
Assume that a model response $y \in \mathbb{R}$ is generated by the function $y = f(\bm{x}) + \epsilon$ with a corresponding input $\bm{x}\in \mathbb{R}^d$, and $\epsilon$ denotes the noise of the measurement or stochasticity of the model and assumes to follow a normal distribution: $\mathcal{N} (0,\sigma_n^2)$. A Gaussian process can be defined as a collection of random variables and any finite number of the random variables follows joint Gaussian distribution \cite{Rasmussen2004}:
\begin{equation}
    f(\bm{x}) \sim GP( m(\bm{x}) , k(\bm{x},\bm{x}^{\prime} ) ),
\end{equation}
where $m(\bm{x})$ is mean function and $k(\bm{x},\bm{x}^{\prime} )$ denotes covariance functions or kernel functions. Generally the the mean function is set to be zero for simplicity and later will be updated during prediction. The kernel functions specify the how the random variables are correlated to each other and also imply the smoothness of the functions. One of the common choice is the radial basis function kernel with automatic relevant determination (ARD) \cite{Rasmussen2004}:
\begin{equation}
    k \left(\bm{x}, \bm{x}^{\prime}\right)=\sigma_{f}^{2} \exp \left(-\frac{1}{2} \sum_{i=1}^{d} \frac{\left(x_{i}-x_{i}^{\prime}\right)^{2}}{\ell_{i}^{2}}\right) ,
\end{equation}
where $\sigma_f^2$ is the signal variance and $\ell_i$ denotes the lengthscale for each input dimension. For a regression problem, an independent Gaussian kernel with variance $\sigma_n^2$ is used to specify the noise in the function. These hyperparameters in the kernel will be determined via the optimization of likelihood function with observed data collection $ (\mathbf{X},\mathbf{y}) = \{(\bm{x}_i,y_i)\}^{N}_{i=1} $:
\begin{equation}
\begin{aligned}
    \arg \max _{\theta} \log p(\mathbf{y}| \mathbf{X}, \boldsymbol{\theta}) = \arg \max _{\theta} \Big[ -\frac{1}{2} \mathbf{y}^{\top}\left( \bm{K} +\sigma_{n}^{2} \mathbf{I}\right)^{-1} \mathbf{y}\\ -\frac{1}{2} \log \left| \bm{K}+\sigma_{n}^{2}\mathbf{I}\right|-\frac{n}{2} \log 2 \pi  \Big],
\end{aligned}
\end{equation}
where $\theta = \{ \sigma_f,\sigma_n,\ell_1,\dots,\ell_d \}$ and $\bm{K} = k(\mathbf{X},\mathbf{X})$. To predict model response at an unevaluated location $\bm{x}^*$, the Gaussian process prior can be rewritten into:
\begin{equation}
    \left[\begin{array}{c} \mathbf{y} \\ f(\bm{x^*}) \end{array}\right] 
\sim 
\mathcal{N}\left(\mathbf{0},\left[\begin{array}{cc}
k( \mathbf{X}, \mathbf{X})+\sigma_{n}^{2} \mathbf{I} & k\left(\mathbf{X},\bm{x}_{*}\right) \\
k\left(\bm{x}^{*}, \mathbf{X}\right) & k\left(\bm{x}^{*}, \bm{x}^{*}\right)
\end{array}\right]\right)
\end{equation}
Conditioning on the observed data, the predictive distribution of the new point $\bm{x^*}$ also follows a normal distribution:
\begin{equation}
    f(\bm{x}^*) | \mathbf{X},\mathbf{y},\bm{x^*} \sim \mathcal{N}\big(\bar{y}^*,\text{Var}(y^*)\big),
\end{equation}
where
\begin{equation*}
\begin{aligned}
    \bar{y}^* &= k(\bm{x}^*,\mathbf{X})[k(\mathbf{X},\mathbf{X})+\sigma_n^2 \mathbf{I}]^{-1}\mathbf{y},\\
    \text{Var}(y^*) &=k(\bm{x}^*,\bm{x}^*) -k(\bm{x}^*,\mathbf{X}) \big[ k(\mathbf{X},\mathbf{X})+\sigma_n^2\mathbf{I}\big]^{-1} k(\mathbf{X,\bm{x}^*}).
\end{aligned}
\end{equation*}
The $\bar{y}^*$ stands for the mean prediction of the response and $\text{Var}(y^*)$ is the predictive variance indicating the uncertainty of the prediction. 

Generally, the Gaussian process regression is applied as a surrogate model to infer the latent function between uncertain inputs and QoIs. However after the decomposition of the model response by SVD, both evaluated and unevaluated model responses can be represented by the projection coefficients on the chosen orthogonal bases, therefore the Gaussian process is now used to learn the mapping between uncertain inputs and projection coefficients of POD and predicts the new coefficients for unevaluated points. The details of the procedure are shown in Algorithm~\ref{alg:GP_POD}.

\begin{algorithm}[tb]
\caption{Constructing a surrogate model for ISR3D with GP and POD}\label{alg:GP_POD}
\vspace{0.08cm}
\textbf{Training}\\
\text{1. Evaluate} $N_{train}$ number of samples using ISR3D and collect the training data $\big\{(\bm{x}_i,\bm{y}_i)\big\}^{N_{train}}_{i=1}$ \\
\text{2. Construct} the snapshot matrix $[\bm{y}_1|\bm{y}_2|\cdots|\bm{y_{N_s}}]$ and perform SVD to obtain $k$ orthogonal bases  $\bm{\Tilde{\Phi}}$ based on the relative energy threshold. \\
\text{3. Project} the output of the training data to each basis and compute the projection coefficients: $\{(\bm{x}_i,\bm{y}_i)\}^{N_{train}}_{i=1}\xrightarrow{POD}\big\{ \big(\bm{x}_i,\alpha_j(\bm{x}_i)\big)\big\}^{N_{areatrain}}_{i=1}  $, where $j=1,\cdots,k$. \\
\text{4. Train} $j$-th single-output GP with uncertain inputs and projection coefficients $\big\{ \big(\bm{x}_i, \alpha_j{(\bm{x}_i)}\big)\big\}$ \vspace{0.05cm}\\
\textbf{Prediction}\\
\text{1. For} an unevaluated point $\bm{x^*}$, use GPs to predict its projection coefficients $\alpha_j(\bm{x^*})$, where $j=1,\cdots,k$.\\
\text{2. Reconstruct} the corresponding model response $\bm{y}^* = \sum_{j=1}^{k} \alpha_j(\bm{x^*})\phi_j$
\end{algorithm}

% inputs to model responses er time $\bm{y} = f(\bm{x})$, where $\bm{y} \in \mathbb{R}^t $ and $\bm{x} \in \mathbb{R}^d$. $t$ and $d$ denote the dimensions of the uncertain inputs and of the model response, respectively. With POD described before, the model response can be decomposed into a weighted summation of multiple orthogonal basis vectors, $f(\bm{x}) \approx  \sum^{k^*}_{i=1} \alpha_i (\bm{x}) \bm{u_i}$. Hence, the outputs are now the projection coefficients $\alpha_i (\bm{x})$ of each basis given the uncertain parameters:
% \begin{equation}
%      \alpha_i = g_i (\bm{x}) + \epsilon_{i}, \; i = 1,2,...,k^*.
% \end{equation}
% Such regression problem can be resolved by Gaussian process method \cite{Rasmussen2004}.

\section{Uncertainty quantification}\label{sec:UQ}
\subsection{Uncertain parameters}
The four epistemic uncertainties considered in the forward uncertainty propagation of ISR3D include endothelium regeneration time, blood flow velocity, the threshold strain for smooth muscle cells bond breaking, and the percentage of fenestration in the internal elastic lamina. Note that all the uncertain parameters except the blood flow velocity are parameters of the SMC submodel.

\textbf{Endothelium regeneration time}: The endothelium regeneration starts right after the denudation caused by the balloon dilation and stent deployment. With sufficiently high wall shear stress from blood flow, the endothelium releases nitric oxide, which behaves as the inhibitor of the proliferation of SMCs. Therefore the rate of endothelial regrowth significantly influences the growth of neointima. In the ISR3D, the regeneration of endothelium cells is modelled to increase linearly up to a coverage of 59\% after 3 days, followed by a full recovery to 100\% after a certain number of days given by the uncertain input \cite{ISR3D2017}. This setting is based on experimental results from Nakazawa et al. \cite{Gaku2010}. However, the exact time for re-endothelialization may vary with many factors, such as the severity of vessel injury, the types of stenting and the degrees of inflammatory response \cite{Teruo2011}. In order to study this uncertain parameter, we consider an average endothelium regeneration time of 15 days and ranges $\pm 50\%$ in the uncertainty quantification.

\textbf{Threshold relative strain}:
The threshold strain is the maximum strain that can be obtained before the bonds between SMCs break. Generally, during the stenting process, the vessel wall is overstretched in the circumferential direction, and therefore the connections between the SMCs (e.g. collagen fibers) are possibly broken and cause microfractures in the tissue. These microfractures may cause inflammation and contribute to the proliferation of SMCs after stenting.

Our choice on the uncertainty of the breaking strain is inferred from stretching experiments \cite{Holzapfel2005,Holzapfel2004} in which the mechanical responses of the coronary arteries under stretch condition were gauged. The result demonstrated that the first intimal rupture occurred at around 110\% strain, and the strain-stress curve became non-smooth when strain reached approximately 120\% . Therefore, we consider the threshold strain around the experimental rupture value 1.1 with an uncertainty of +/-20\% in our UQ experiment. 
Note that the measurements in \cite{Holzapfel2005,Holzapfel2004} started from an unstrained sample, while in our model the vessel is pre-strained 30\% due to being pressurized by the flowing blood inside it.

\textbf{Blood flow velocity}:
Blood flow, as one of the mechanical factors, also plays an important role in the growth of neointima \cite{Kohler1992,CALIFF19912}. High enough wall shear stress in the vessel accelerates the production of nitric oxide in endothelial cells, which acts as an inhibitor of SMCs proliferation. 

As mentioned before, the blood flow in the simulation is modelled as a steady flow with a constant parabolic inlet boundary condition. The velocity data from \cite{Huo2007} was applied to compute time-averaged velocity and converted to the parabolic profiles, the maximum velocity of which is $0.266 \text{m/s}$. Due to the measurement error and potential variety of velocity for individual vessels, we presume a large uncertainty in the data and vary 50\% based on the average values $0.266 \text{m/s}$.

\textbf{Fenestration percentage}:
The internal elastic lamina is modelled in ISR3D as a layer of agents on the inner surface of the vessel wall \cite{ISR3D2017}. The fenestrations on IEL significantly affect the initial growth of SMCs as they allow SMCs to migrate into the blood vessel and start proliferating there. However, the SMCs in ISR3D are not able to change shape to migrate through the fenestrations, unlike real SMCs. Therefore, a certain percentage of IEL agents is switched to SMCs in ISR3D, to obtain a smaller amount of very large fenestrations, with the same total surface area as in the experiment. The uncertainty ranges for this parameter are obtained from \cite{KWON1998283} where the percentage of fenestration in the hypercholesterolemic group is approximately $7.5\%$ and in the control group is approximately $3.5\%$. To include and study the scenarios for both cases, we consider the parameters to vary from $2\%$ to $10\%$. 

The ranges of all the uncertain parameters mentioned above are given in Table~\ref{tab:UncertainPara} and the distributions of the uncertainties are all assumed to be uniform. 

\begin{table}[tb]
\centering
\begin{adjustbox}{width=0.7\textwidth}
\begin{tabular}{c|c|c|c|c}
\toprule
\textbf{Uncertain Parameters}       & \textbf{Ranges (Min)} & \textbf{Ranges (Max)} & \textbf{Unit} & \textbf{CV} \\ 
\midrule
endothelium regeneration   & 10           & 20           & day      & 0.19     \\ 
blood flow velocity        & 0.133          & 0.399          & m/s    & 0.29        \\ 
relative threshold strain          & 0.446          & 0.785          & /    & 0.16         \\ 
percentage of fenestration & 2            & 10           & \%      & 0.38      \\ \bottomrule
\end{tabular}
\end{adjustbox}
\caption{Ranges, units and coefficient of variation (CV) of uncertain parameters of ISR3D model. Note that the relative threshold strain is calculated with 30\% pre-strained. Note that the measurements in \cite{Holzapfel2005,Holzapfel2004} started from an unstrained sample while in our model the vessel is pre-strained 30\%. Therefore, strain listed by Holzapfel et al, $\sigma_{\text{abso}}$ is scaled to obtain the relative deformation of our pre-strained tissue by $\sigma_{\text{rela}}= \frac{\sigma_{\text{abso}}+1}{1.3}-1$).}
\label{tab:UncertainPara}
\end{table}
% describe four uncertain parameters

\subsection{Uncertainty estimations and sensitivity analysis}\label{subsec:MFMC}
For the UQ we applied the quasi-Monte Carlo (qMC) sampling method with Sobol sequence \cite{SOBOL199055}. The method allows the sample to be more evenly distributed in the domain which leads to a better convergence rate compared to the standard random sampling.

To investigate the uncertainty propagation of the uncertain inputs through the model, mean, variance, probability density function, and coefficient of variation are estimated. Besides, global sensitivity analysis has been conducted to study how much each uncertain input has contributed to the uncertainty of QoIs. The variance-based method (Sobol method) \cite{SALTELLI2010259} is applied, which assumes that the latent functions $f(x)$ can be decomposed into a combination of functions of individual uncertain inputs and their higher-order interactions, which also leads to the following decomposition of the variance \cite{SALTELLI2010259}:
\begin{equation}
    \text{Var}f(x)=\sum_{i} V_{i}+\sum_{i} \sum_{j>i} V_{i j}+\cdots+V_{12 \ldots d},
\end{equation}
where $V_i$, $V_{ij}$, $V_{12\dots d}$  stand for the partial variance contributed by $i$-th uncertain input, by the interactions between $i$-th and $j$-th uncertain inputs and by higher-order interactions. The first order Sobol indices indicate the independent contributions from the partial variance of each single uncertain input:
\begin{equation}
    S_{i}=\frac{\operatorname{Var}_{x_{i}}}{\operatorname{Var} f( {\bm{x} )}} =\frac{ \operatorname{Var}_{x_{i}}\left(E_{\bm{x}_{\sim i}}\left( f(\bm{x}) \mid x_{i}\right)\right)}{\operatorname{Var} f(\bm{x})}
\end{equation}
where $\bm{x}_{\sim i}$ denotes a vector of all uncertain paramters in $\bm{x}$ except $x_i$. The total sensitivity indices take all the relevant contributions of a uncertain input into account: 
\begin{equation}
    S_{T_i}=\frac{\operatorname{Var}^{\text{total}}_{x_{i}}}{\operatorname{Var}f( {\bm{x} )}}
    =1-\frac{\operatorname{Var}_{\bm{x}_{\sim i}}\left(E_{x_{i}}\left( f(\bm{x}) \mid \bm{x}_{\sim i}\right)\right)}{\operatorname{Var} f( \bm{x})}
\end{equation}
All the sensitivity indices mentioned above are computed by Saltelli's method \cite{SALTELLI2010259}.

\section{Results}\label{sec:result}
To train the surrogate models, 512 samples were generated by the qMC method and evaluated by the ISR3D model. Before the surrogate model was deployed to the UQ experiment, the surrogates were validated with four-fold cross-validation. We measured the approximation error of both POD and GP regression with the relative $L^2$ norm:
\begin{equation}
    e_{POD} = \sum_{i=1}^{N_{cv}} \sqrt{\frac{\lVert \bm{y}_i -\bm{\Tilde{\Phi}} \bm{\Tilde{\Phi}}^T \bm{y}_i \rVert_{L^2}}{\lVert \bm{y}_i \rVert_{L^2}}},\hspace{1cm}
    e_{GP} = \sum_{i=1}^{N_{cv}} \sqrt{\frac{\lVert \bm{y}_i - \bm{\Tilde{\Phi}}\bm{\alpha}(\bm{x}_i) \rVert_{L^2}}{\lVert \bm{y}_i \rVert_{L^2}}}.
\end{equation}

\begin{figure}[tb]
\centering
\begin{subfigure}{.5\textwidth}
  \raggedright
  \includegraphics[width=0.88\linewidth]{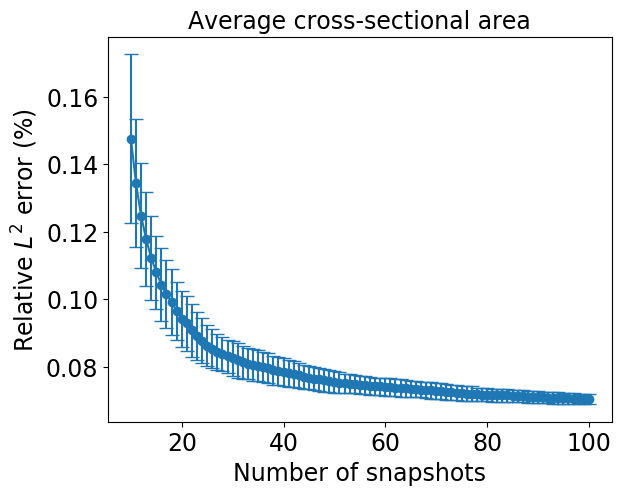}
  \caption{}
  \label{fig:POD_AA}
\end{subfigure}%
\begin{subfigure}{.5\textwidth}
  \raggedright
  \includegraphics[width=0.87\linewidth]{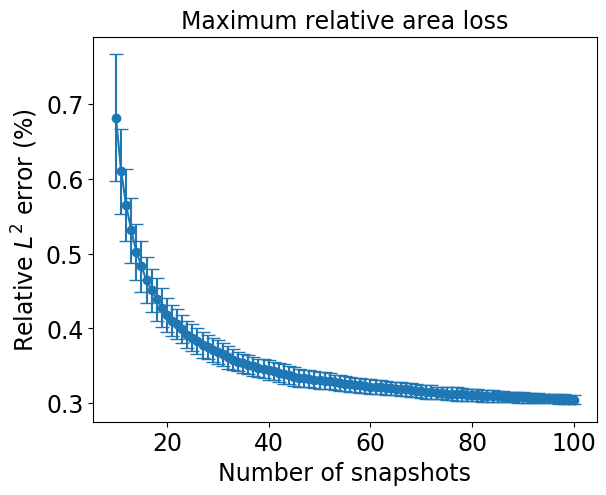}
  \caption{}
  \label{fig:POD_RA}
\end{subfigure}
\caption{Relative $L^2$ error of POD with different number snapshot of two QoIs, average cross-sectional area and maximum relative area loss using cross-validation. The error bar stands for the standard deviation computed from 100 replications}
\label{fig:PODerror}
\end{figure}

\noindent In the cross-validation of POD, a certain number of snapshots were randomly taken from the training dataset and constructed the snapshot matrix for SVD. The validation dataset was used to measure the approximation error. The relative $L^2$ error of POD approximation with a different number of snapshots of both QoIs are shown in Figure~\ref{fig:PODerror}. The average relative $L^2$ error gradually decreases to around $0.07\%$ and $0.3\%$ respectively with the number of snapshots reaching 100. The tendency of the curve shows that the error has almost converged to a limit; a further increase in the number of snapshots will not greatly improve performance. The low standard deviation of the error shows that there is no significant influence on the choice snapshots. Therefore, we randomly chose 100 snapshots from the output of the training data for the POD in the construction of the surrogate model. 

\begin{figure}[tb!]
\centering
\begin{subfigure}{.5\textwidth}
\hspace{0.05cm}
  \includegraphics[width=0.9\linewidth]{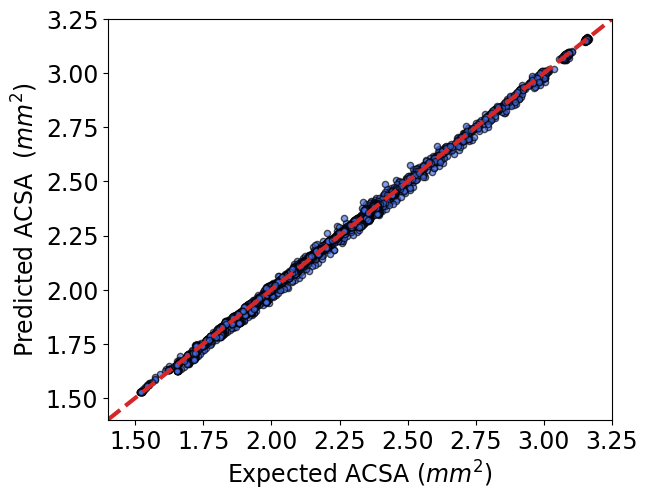}
  \caption{}
  \label{fig:POD_AA}
\end{subfigure}%
\begin{subfigure}{.5\textwidth}
%   \raggedright
  \hspace{0.1cm}
  \includegraphics[width=0.86\linewidth]{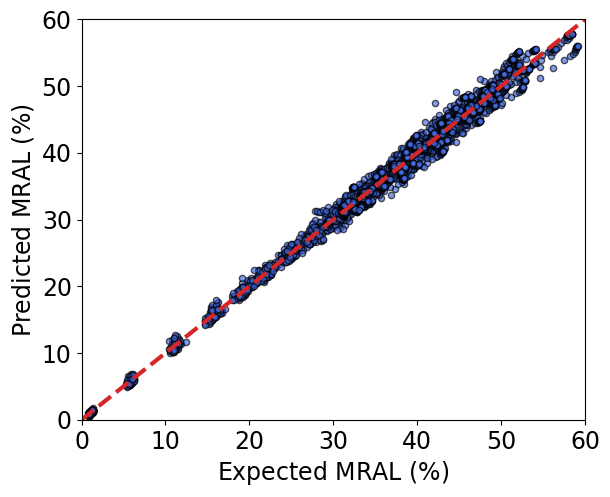}
  \caption{}
  \label{fig:POD_RA}
\end{subfigure}
\caption{A comparison between expected QoIs and predicted QoIs over all time steps from a cross-validation, where the diagonal lines denote the precise predictions of expected values. ACSA denotes average cross-sectional area and MRAL denotes maximum relative area loss.}
\label{fig:GP}
\end{figure}

To test the performance of the GP regression, another four-fold cross-validation was performed with 100 repetitions. The predicted projection coefficients were first used to reconstruct their original model responses and subsequently compared to the expected output from the validation dataset. Comparisons of the predicted QoIs versus expected QoIs over all time steps are demonstrated in Figure~\ref{fig:GP}. The resulting points are clustered around the diagonal line indicating that the GP has inferred the underlying functions well. The average relative $L^2$ error is $0.52\%$ for the average cross-sectional area and $2.52\%$ for the relative maximum area loss.

\begin{figure}[tb!]
\centering
\includegraphics[width=0.95\textwidth]{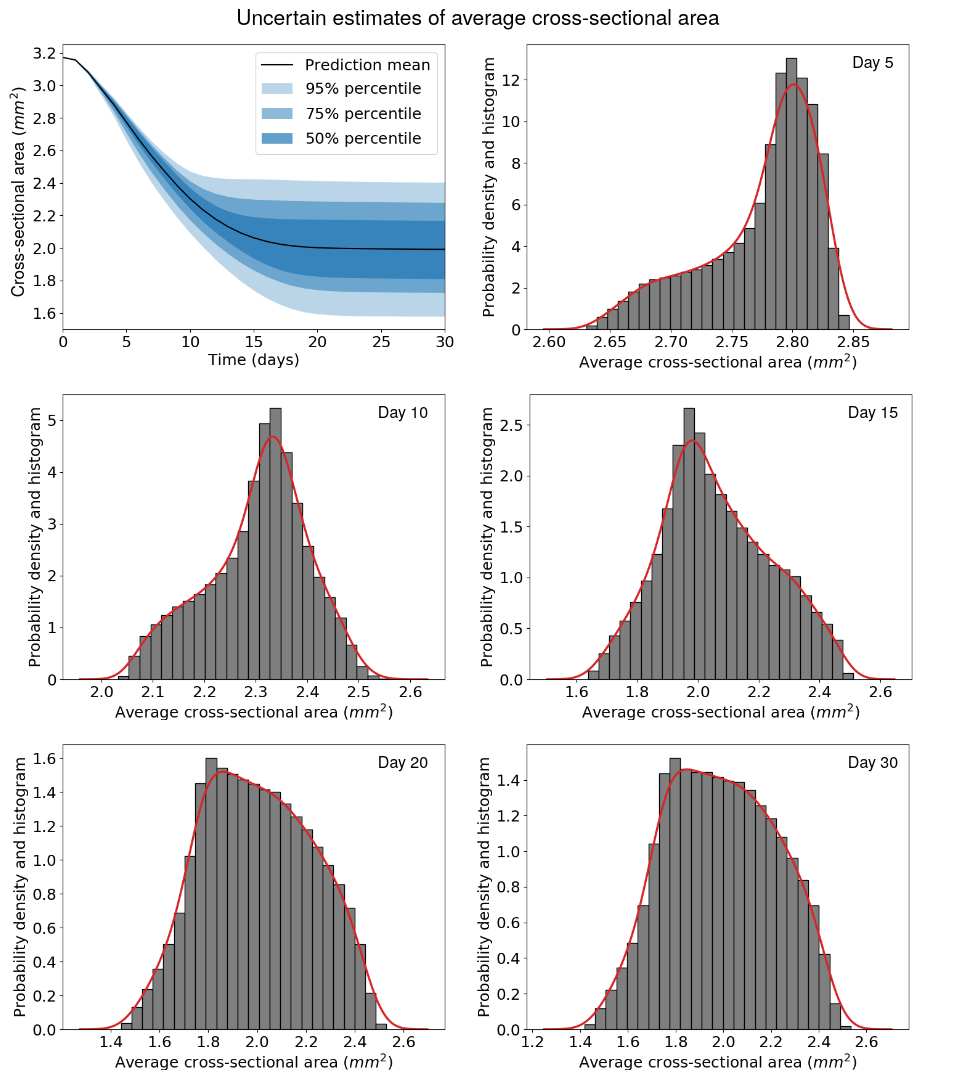}
\caption{Mean, 50\%, 75\% and 95\% percentile of the average cross-sectional area of the lumen over time with quasi-Monte Carlo method and corresponding histogram and probability density function at day 5, 10, 15, 20 and 30.}
\label{fig:UQAA}
\end{figure}

After the validations of surrogate models, the UQ experiments for both QoIs were performed. We applied qMC method to draw $10^5$ samples from the uncertain input domain and fed to the surrogate models. The mean and 50\%, 75\% and 95\% percentile estimations of average cross-sectional area over time are shown in Figure~\ref{fig:UQAA}. The corresponding histogram and probability density functions (PDF) of day 5, 10, 15, 20, and 30 are also shown in the same figure. The initial average cross-sectional area after stenting was around $3.17~mm^2$. With the evolution of time, the cross-sectional area gradually reduced due to the neointimal growth. The mean estimation of the average cross-sectional area shows that the neointimal growth was slow at the beginning but started to accelerate after day 1. An almost linear growth between day 1 and day 10 was observed followed by a descending growth rate until all the growth stopped at around day 22. The upper boundary of the 95\% percentile shows that some samples stopped growing shortly after day 10 due to the short re-endothelialization time, while a few other cases did not stop before 22 days.

\begin{figure}[tb!]
\centering
\includegraphics[width=0.92\textwidth]{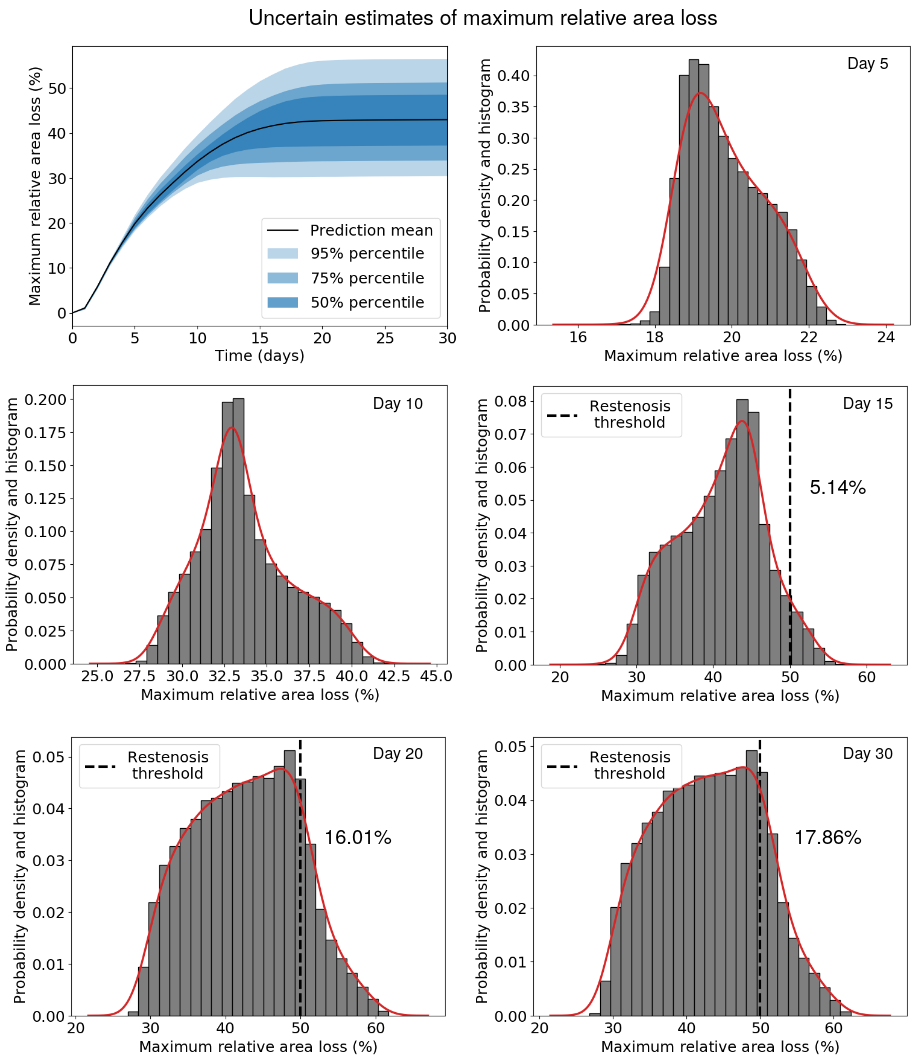}
\caption{Mean, 50\%, 75\% and 95\% percentile of the maximum relative area loss over time with quasi-Monte Carlo method and corresponding histogram and probability density function at day 5, 10, 15, 20 and 30.}
\label{fig:UQRA}
\end{figure}

The PDFs and histograms in Figure~\ref{fig:UQAA} show the details of the distributions of day 5, 10, 15, 20, and 30. On day 5, most of the samples cluster around 2.8 $mm^2$ and a small part of the samples have a lower average cross-sectional area up to 2.63 $mm^2$. A certain number of samples already stopped growing between day 10 and 20. The early stop usually means a small amount of neointima and contributes to the right tail of the distributions (around 2.4 $mm^2$ to 2.6 $mm^2$), while the rest of the samples still shifted to the left due to the growth. The difference between day 20 and day 30 is minor, indicating that the growth in most of the samples had stopped before day 20.  

Similar patterns can be observed for the maximum relative area loss in Figure~\ref{fig:UQRA}. The distribution at day 30 shows that most of the simulations ended up with $30\%$ to $60\%$ area loss. Assuming that the restenosis happened when the area loss reaches $50\%$, about $5\%$, $16\%$, $18\%$ of the simulations had reached the restenosis threshold at day 15, 20 and 30, respectively.

\begin{table}[tb!]
\centering
\begin{adjustbox}{width=0.9\textwidth}
\begin{tabular}{c|c|c|c|c|c|c|c|c|c|c}
\toprule
                    & \multicolumn{5}{c|}{Average cross-sectional area ($mm^2$)} & \multicolumn{5}{c}{Maximum relative area loss (\%)} \\ \midrule
Estimates & day 5   & day 10   & day 15   & day 20  & day 30  & day 5  & day 10  & day 15  & day 20  & day 30  \\ 
\midrule
Mean & 2.774 & 2.303 & 2.062 & 2.002 & 1.991 & 19.848 & 33.674       & 40.962 & 42.721 & 42.958 \\
SD  & 0.046 & 0.098 & 0.179 & 0.224 & 0.226 & 1.023  & 2.745       & 5.672  & 7.037 & 7.124 \\
CV   & 1.658\% & 4.255\% & 8.681\% & 11.189\% & 11.351\% & 5.154\%  & 8.152\% & 13.847\%  & 16.472\% & 16.591\% \\
Restenosis & / & / & / & /& / &  0\% &  0\% & 5.123\% & 16.047\% & 17.873\% \\
\bottomrule
\end{tabular}
\end{adjustbox}
\caption{Mean, standard deviation (SD), CV (in percentage) and percentage of restenosis at day 5, 10 ,15, 20 and 30 for both QoIs computed from 100 repetitions of the UQ estimation. Due to the large number of samples used, the confidence interval of the evaluations for each estimates is extremely small ($\leq 10^{-5}$).}
\label{tab:mean}
\end{table}

Table~\ref{tab:mean} provides detailed information of mean, standard deviation (SD), CV at day 5, 10, 15, 20 and 30 of both QoIs computed by 100 replications of the UQ experiment. Around 11.3\% and 16.6\% of uncertainty are observed from the average cross-sectional area and maximum relative area loss respectively.

\begin{figure}[tb!]
\centering
\includegraphics[width=0.98\textwidth]{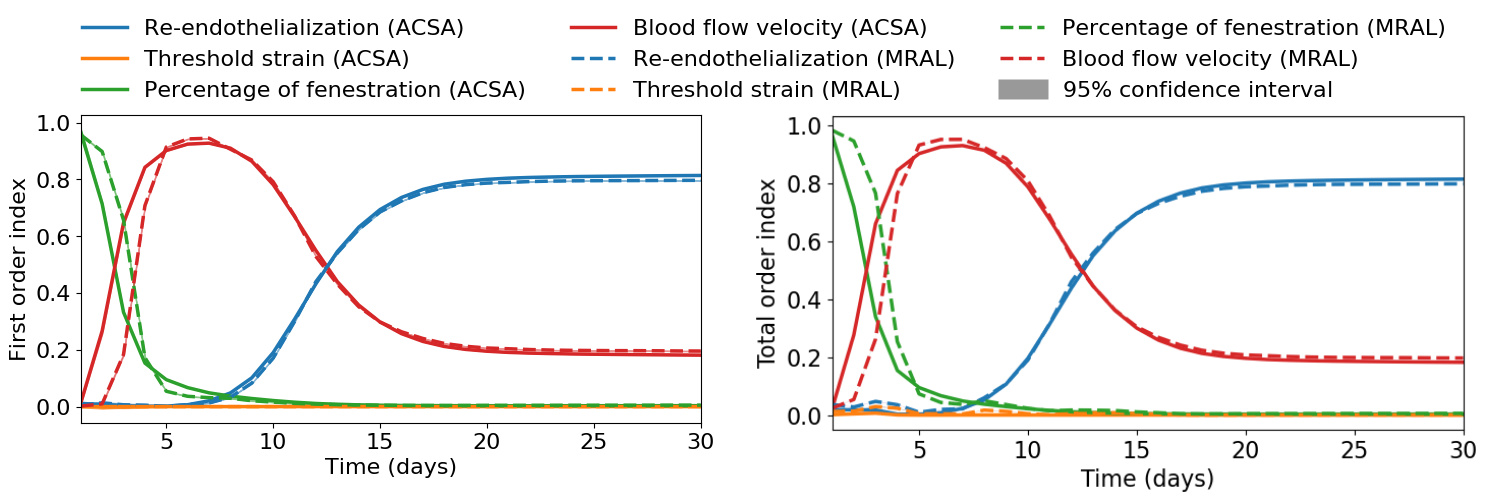}
\caption{First order and total Sobol sensitivity indices of both QoIs (ACSA - average cross-sectional area, MRAL - maximum relative area loss) and corresponding 95\% confidence interval of each estimate based on 100 replica computing.}
\label{fig:SAFO}
\end{figure}

Apart from the uncertainty estimations, sensitivity analysis has also been performed. The sensitivity analysis was performed with $5 \times 10^5$ samples using Sobol sequence and was repeated 100 times to compute the confidence interval. The first order indices of the four uncertain inputs over time for both QoIs are shown in Figure~\ref{fig:SAFO}. The confidence interval can hardly be seen in the figure, indicating extremely small uncertainty in our sensitivity estimations. For both QoIs, a dominant influence of the fenestration percentage can be observed at the initial stage of the process and keeps decreasing over time. It has almost no impact after 10 days. The blood flow velocity is a critical factor on the growth throughout the entire process and shows significant influences in between day 5 to day 10, and gradually falls to around 0.2, while the endothelium regeneration times shows an increasing effect and plays the most important role after 13 days. The threshold strain is relatively not important compared to the other uncertain inputs. The total order indices of both QoIs are very similar to their first order result meaning that there is little higher-order interaction between the uncertain inputs.

\begin{figure}[tb!]
\centering
\includegraphics[width=0.98\textwidth]{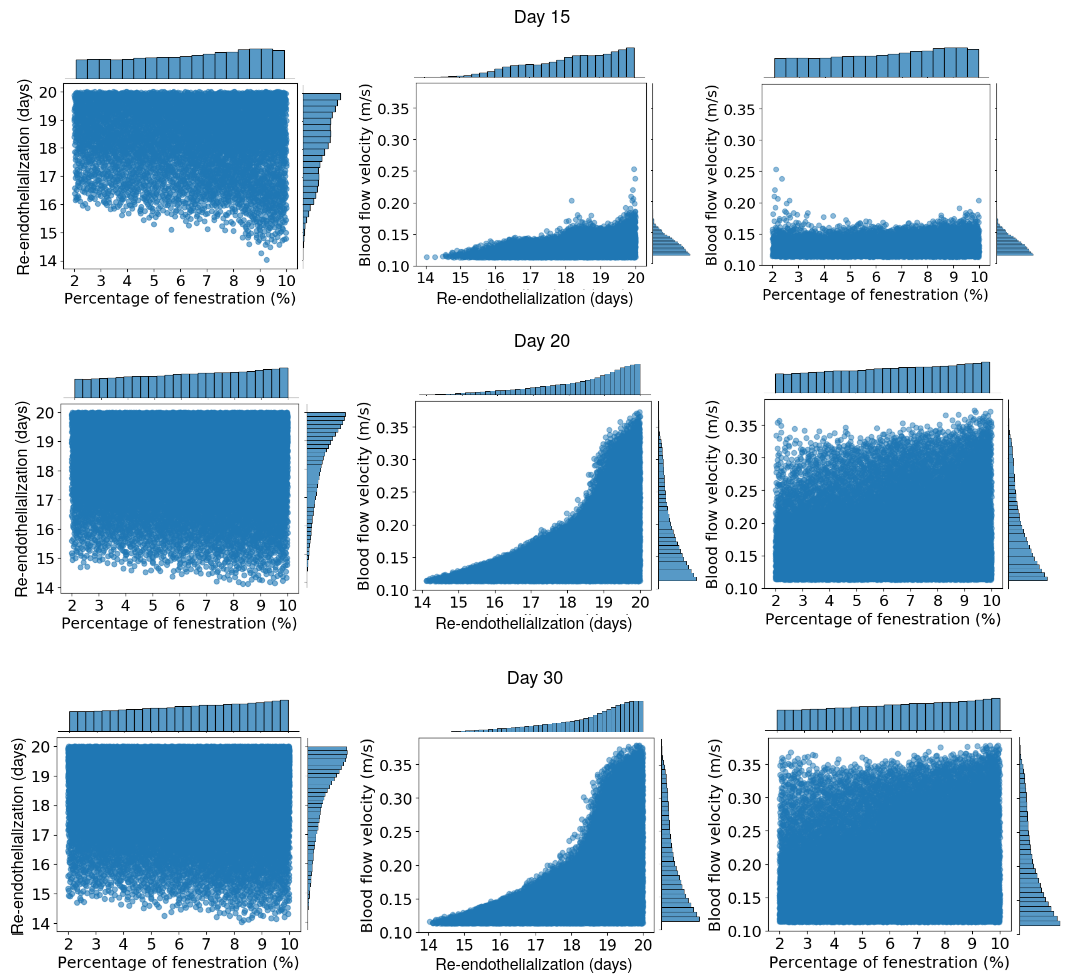}
\caption{Restenosis sample scatter distributions and corresponding histograms at day 15, 20 and 30.}
\label{fig:Samples}
\end{figure}

To further investigate the relations between uncertain inputs and restenosis, scatter distributions and histograms of the samples which reached the restenosis threshold at day 15, 20 and 30 are shown in Figure~\ref{fig:Samples}. Note the threshold strain is not shown in the Figure since the sensitivity analysis result suggested that it is not important in the process. 

In the left column of Figure~\ref{fig:Samples}, scatter distributions of samples in terms of fenestration percentage and re-endothelialization time are shown. The range of re-endothelialization time falls between days 14 to 20, meanwhile a clear degression tendency can be observed from the corresponding histogram. The range of fenestration percentage shows that the restenosis can happen even with the lowest fenestration percentage, but the probability decreases slightly as the percentage drops. The middle column demonstrates the scatter distributions of re-endothelialization time and blood flow velocity. At day 15, only the cases with rather low blood flow velocity (under 0.27 $m/s$) reached the restenosis threshold. However at the end of the simulations, the upper bound rose to 0.38 $m/s$. Unlike the left and right columns, a clear separation can be found between restenotic samples and the rest of the domain. The right column is based on fenestration percentage and blood flow velocity. Same patterns could be observed. The influence of fenestration percentage is rather minor while the value of blood flow velocity significantly affects the possibility of restenosis.

The speedup of the entire UQ experiment using the surrogate model has also been estimated. Table~\ref{tab:speedup} shows the details of the computational cost including the average core hour for model evaluation with ISR3D and surrogate model $\mathcal{T}_{\text{ISR}}$, training data generation $\mathcal{T}_{\text{sample}}$, and surrogate training $\mathcal{T}_{\text{train}}$. Both training and prediction of a surrogate model was extremely fast. The most computational expensive part was the generation of training data with ISR3D. The average core hour for each evaluation was around 585. Since in this case $N_{\text{UQ}}\mathcal{T}_{\text{ISR}} + \mathcal{T}_{\text{train}}$ is negligible as compared to $\mathcal{T}_{\text{train}}$, we find that the speedup equals $N_{\text{UQ}}/512 \approx 976.6$.

\begin{table}[tb!]
\begin{adjustbox}{width=\textwidth}
\begin{tabular}{cccccc}
\toprule
UQ method & $\mathcal{T}_{\text{ISR}}$ (core hour) & $\mathcal{T}_{\text{train}}$ (core hour) & $\mathcal{T}_{\text{sample}}$ (core hour) & $N_{\text{UQ}}$                 & Speedup of UQ \\
\midrule
qMC       & 585.1                  & /             & /                      & / & 1       \\
qMC (surrogate)        & $6.1\times 10^{-8}$              & $3.1\times 10^{-2}$          & $585.1\times 512$              & $5\times 10^5$ & 976.6 \\
\bottomrule
\end{tabular}
\end{adjustbox}
\caption{The computation cost of ISR3D model and surrogate model. $\mathcal{T_{\text{ISR}}}$ denotes the core hour to finish one single run of the simulation. Each simulation was performed exclusively on a node with $2\times12$-core 2.6 GHz Intel Xeon E5-2690 v3 (Haswell) CPUs of Dutch supercomputer \textit{Cartesius}. $\mathcal{T}_{\text{train}}$ and $\mathcal{T}_{\text{sample}}$ stand for the sample generation time and training time for the surrogate model. $N_{\text{UQ}}$ denotes the number of samples used in a UQ experiment}
\label{tab:speedup}
\end{table}

\section{Discussion}\label{sec:dis}
The result of surrogate modeling shows that the combination of POD and Gaussian process regression performs well. The decomposition and reconstruction of the model response with POD save the computational effort for regression and provide a convenient and consistent way to cover the entire model response over time. In this work, the snapshot matrix was constructed by 100 randomly- chosen snapshots from the training dataset generated by quasi MC sampling. Adaptive sampling method \cite{LU2019264} can be used to choose more representative snapshots with error estimations, however it is unnecessary as a relatively large training dataset was available and the approximation error could be properly controlled.  

The Gaussian process was then applied to infer the latent functions between uncertain inputs and projection coefficients of POD. In the cross-validation of the surrogate model, the relative $L^2$ error of the maximum relative area loss is slightly larger than the other QoI. It is mainly due to its way of computing relative area loss which required a division of the initial cross-sectional area. The initial cross-sectional areas at each slice of the lumen are different and thus introduced the noise into the data. Therefore the regression performance of such QoI was slightly worse than the others. 

For the uncertainty quantification, around 11\% and 16\% of uncertainty are observed from the average cross-sectional area and maximum relative area loss respectively. The uncertainties in the output are mainly contributed by fenestration percentage, blood flow velocity and endothelium recovery time. The fenestration percentage is important at the beginning because a larger amount of fenestrations allows more SMCs to migrate to the vessel lumen and proliferate. However such impact drops sharply to almost 0 in the first 5 days, as the SMCs form a continuous layer over the IEL. Meanwhile the blood flow velocity starts to dominate the variance between day 5 and day 10. During day 5, re-endothelialization coverage varied from 63\% to 67\% and raised up to 73\% to 87\% by day 10 which means that if the wall shear stress is sufficiently high, a large percentage of cells at the lumen surface could already have their growth inhibited by nitric oxide. After day 10, the influence of the blood flow velocity drops gradually and is replaced by re-endothelialization. Figure~\ref{fig:Samples} shows that at the end of the simulations, the influences of fenestration percentage is relatively minor compared to the effect of blood flow velocity and endothelium regeneration time. It suggests that the scenarios with a high fenestration percentage, such as hypercholesterolemia, might not have a high impact on restenosis probability if other parameters such as endothelium regeneration time can be strictly controlled.

In this work, we studied four biological uncertain parameters. We quantified their uncertainty propagation and sensitivity for two QoIs adapted for in-silico models from clinically recognized metrics. This helps us to better understand the underlying contribution of these parameters to restenosis. In addition to the investigated biological factors, other factors and scenarios can be also studied via ISR3D, for example, variability in the stenting procedure, such as deployment depth or malapposition of the stent. Through the UQ analysis, the potential effect of such factors can be quantified and studied. Additionally, different scenarios, such as small/large vessel diameters and the tortuosity of the stented vessel, can also significantly influence the outcome of a simulation. We leave the study of these factors, which all affect the initial shape of the stented vessel, to our future work.  

The ISR3D model itself has several limitations. First, it does not account for the inflammation processes, which are important during the early stages of post-stenting. Second, the geometry used in the UQ experiment is not based on any particular vessel, and instead is a piece of a perfectly cylindrical tube, and the stent fits the curvature of the vessel perfectly and is radially expanded in a uniform way. All these factors may contribute to the underestimation of restenotic growth. For example, Morton et al. \cite{Morton324} reported the area loss of 62\% for porcine vessels of a similar diameter and deployment depth with NIR stent, which is very close to the upper bound of the distribution predicted by the model. Nevertheless, the experimental values lie within the distribution, further confirming that the ranges selected for UQ reasonably overlap with the physiological ranges. There are also other limitations in the model we use, discussed in detail in \cite{ISR3D2017,Zun2019}.

\section{Conclusion}\label{sec:con}
The uncertainty quantification and sensitivity analysis of a multiscale model ISR3D was performed. The uncertainty propagation from four parameters: endothelium regeneration time, threshold strain, percentage of fenestration and blood flow velocity; to two QoIs: average cross-sectional area and relative maximum area loss; are investigated. Due to the high computational cost of ISR3D, surrogate modelling techniques were applied. The QoIs over time were, first, decomposed by proper orthogonal decomposition and the resulting projection coefficients were learned by a Gaussian process regression model. Cross-validations are applied to validate the performance of the surrogate model. The surrogate model was subsequently deployed in the UQ experiment to replace the original model. The uncertainty quantification and sensitivity analysis results showed that the blood flow velocity and endothelium regeneration time have significant influence on the neointima growth and result in restenosis, while the impact from fenestration percentage is limited and the threshold strain barely has any influence on the process.

\section{Funding}
This project has received funding from the European Union Horizon 2020 research and innovation programme under grant agreements \#800925 (VECMA project), \#777119 (InSilc project), \#101016503 (In Silico World project). PZ acknowledges funding from the Russian Science Foundation under agreement \#20-71-10108. This work was carried out on the Dutch national e-infrastructure, with the support of SURF Cooperative and financial support from the Nederlandse Organisatie voor Wetenschappelijk Onderzoek (Netherlands Organization for Science Research, NWO).



%% References



%% New version of the num-names style
% \bibliographystyle{abbrv}
\bibliographystyle{elsarticle-num}

\bibliography{sample.bib}
% \bibliographystyle{elsarticle-num-names}






% \end{thebibliography}

\end{document}